\documentstyle[epsf]{mn}

\newif\ifAMStwofonts
\AMStwofontstrue

\makeatletter
\def\figsize{\ifSFB@referee\epsfxsize=0.5\hsize\else\epsfxsize=\hsize\fi}
\makeatother
\def\n(#1){{}$^{\rm(#1)}$}
\gdef\0{\phantom{0}}
\def\eq#1 {\begin{equation} #1 \end{equation}}
\def\Ref{\bibitem{}}
\def\symbol#1{\hbox{$#1$}}
\def\about{\symbol{\sim}}
\def\tF   {\symbol{\tau_{\rm F}}}
\def\lsca {\symbol{\lambda_{\rm sca}}}
\def\Abol {\symbol{A_{\rm bol}}}
\def\Tstar{\symbol{T_\ast}}
\def\rstar{\symbol{r_\ast}}
\def\alam {\symbol{a_\lambda}}

\def\Lo   {{\hbox{L$_\odot$}}}

\def\Mdot {{\hbox{$\dot M$}}}

\def\mic  {{\hbox{$\umu$m}}}

\ifoldfss
  \ifCUPmtlplainloaded \else
    \NewTextAlphabet{textbfit} {cmbxti10} {}
    \NewTextAlphabet{textbfss} {cmssbx10} {}
    \NewMathAlphabet{mathbfit} {cmbxti10} {} 
    \NewMathAlphabet{mathbfss} {cmssbx10} {} 
  \fi
  \ifAMStwofonts
    \ifCUPmtlplainloaded \else
      \NewSymbolFont{upmath} {eurm10}
      \NewSymbolFont{AMSa} {msam10}
      \NewMathSymbol{\upi}     {0}{upmath}{19}
      \NewMathSymbol{\umu}     {0}{upmath}{16}
      \NewMathSymbol{\upartial}{0}{upmath}{40}
      \NewMathSymbol{\leqslant}{3}{AMSa}{36}
      \NewMathSymbol{\geqslant}{3}{AMSa}{3E}

       \let\le=\leqslant
       \let\ge=\geqslant
    \fi
  \fi
\fi 

\ifnfssone
  \newmathalphabet{\mathit}
  \addtoversion{normal}{\mathit}{cmr}{m}{it}
  \addtoversion{bold}{\mathit}{cmr}{bx}{it}
  \newmathalphabet{\mathbfit} 
  \addtoversion{normal}{\mathbfit}{cmr}{bx}{it}
  \addtoversion{bold}{\mathbfit}{cmr}{bx}{it}
  \newmathalphabet{\mathbfss} 
  \addtoversion{normal}{\mathbfss}{cmss}{bx}{n}
  \addtoversion{bold}{\mathbfss}{cmss}{bx}{n}
  \ifAMStwofonts
    \ifCUPmtlplainloaded \else
      \UseAMStwoboldmath
      \makeatletter
      \new@mathgroup\upmath@group
      \define@mathgroup\mv@normal\upmath@group{eur}{m}{n}
      \define@mathgroup\mv@bold\upmath@group{eur}{b}{n}
      \edef\UPM{\hexnumber\upmath@group}
      \new@mathgroup\amsa@group
      \define@mathgroup\mv@normal\amsa@group{msa}{m}{n}
      \define@mathgroup\mv@bold\amsa@group{msa}{m}{n}
      \edef\AMSa{\hexnumber\amsa@group}
      \makeatother
      \mathchardef\upi="0\UPM19
      \mathchardef\umu="0\UPM16
      \mathchardef\upartial="0\UPM40
      \mathchardef\leqslant="3\AMSa36
      \mathchardef\geqslant="3\AMSa3E

       \let\le=\leqslant
       \let\ge=\geqslant
    \fi
  \fi
\fi 

\ifnfsstwo
  \DeclareMathAlphabet{\mathbfit}{OT1}{cmr}{bx}{it}
  \SetMathAlphabet\mathbfit{bold}{OT1}{cmr}{bx}{it}
  \DeclareMathAlphabet{\mathbfss}{OT1}{cmss}{bx}{n}
  \SetMathAlphabet\mathbfss{bold}{OT1}{cmss}{bx}{n}
  \ifAMStwofonts
    \ifCUPmtlplainloaded \else
      \DeclareSymbolFont{UPM}{U}{eur}{m}{n}
      \SetSymbolFont{UPM}{bold}{U}{eur}{b}{n}
      \DeclareSymbolFont{AMSa}{U}{msa}{m}{n}
      \DeclareMathSymbol{\upi}{0}{UPM}{"19}
      \DeclareMathSymbol{\umu}{0}{UPM}{"16}
      \DeclareMathSymbol{\upartial}{0}{UPM}{"40}
      \DeclareMathSymbol{\leqslant}{3}{AMSa}{"36}
      \DeclareMathSymbol{\geqslant}{3}{AMSa}{"3E}

       \let\le=\leqslant
       \let\ge=\geqslant
    \fi
  \fi
\fi 

\ifCUPmtlplainloaded \else
  \ifAMStwofonts \else 
    \def\upi{\pi}
    \def\umu{\mu}
    \def\upartial{\partial}
  \fi
\fi

\pagerange{\pageref{firstpage}--\pageref{lastpage}}
\pubyear{1996}

\title[Infrared imaging of late-type stars]
                   {Infrared imaging of late-type stars}

\author[\v{Z}. Ivezi\'{c} and M. Elitzur]
                     { \v{Z}eljko Ivezi\'{c} and Moshe Elitzur \\
                        Department of Physics and Astronomy,
                 University of Kentucky, Lexington, KY 40506-0055, USA\\
                   e-mail: ivezic@pa.uky.edu, moshe@pa.uky.edu }

\date{Accepted 1995 November 23. Received 1995 November 23;
      in original form 1995 July 31}

\begin{document}

\maketitle

\label{firstpage}

\begin                             {abstract}
Infrared imaging properties of dusty winds around late-type stars are
investigated in detail, employing a self-consistent model that couples the
equations of motion and radiative transfer.  Because of general scaling
properties, the angular profiles of surface brightness are self-similar. In
any given star, the profile shape is determined essentially by overall
optical depth at each wavelength and it is self-similarly scaled by the size
of the dust condensation zone.  We find that the mid-IR is the best wavelength
range to measure directly the angular size of this zone, and from {\it IRAS}
data we identify the 15 best candidates for such future observations.  We also
show that the visibility function at short wavelengths ($\la$ 2 \mic) directly
determines the scattering optical depth, and produce theoretical visibility
curves for various characteristic wavelengths and the entire parameter range
relevant to late-type stars.  The infrared emission should display time
variability because of cyclical changes in overall optical depth, reflecting
luminosity-induced movement of the dust condensation point. Calculations of the
wavelength dependence of photometric amplitudes and time variability of
envelope sizes are in agreement with observations; envelopes are bigger and
bluer at maximum light.
\end{abstract}

\section                    { INTRODUCTION }

Ongoing progress in imaging techniques, expected to continue at an accelerated
pace (e.g.\ Gezari \& Backman 1993; Skinner et al.\ 1994), enables spatial
resolution of infrared sources.  The wealth of details provided by
high-resolution data requires a commensurate theoretical effort, which can be
expected in general to involve a large number of free parameters. The freedom
afforded by many parameters usually produces good fits but casts doubt on the
uniqueness of such models.

Late-type stars are a notable exception. Recently we have shown that, given
the grain composition, the IR signature of dusty envelopes around such stars
is essentially determined by a single parameter -- the flux-averaged optical
depth \tF\ (Ivezi\'c \& Elitzur 1995; hereafter IE95). This parameter alone
controls both the dynamics and IR emission.  The fine resolution achieved in
recent interferometric observations (envelopes are resolved on angular scales
as small as 0.03 arcsec at \about\ 10 \mic; see Danchi et al.\ 1994) merits
detailed modelling of these stars, which we offer here. The surface brightness
distribution can be calculated once the density and temperature distributions
of the dust in the envelope are known (e.g. Crabtree \& Martin 1979).  In
previous studies, both distributions were usually modelled by power laws
prescribed beforehand. Instead, here we determine these distributions
self-consistently from a complete, detailed model that solves the envelope
dynamics and radiative transfer simultaneously.  This self-consistent approach
enables us to compare different types of observations and to investigate
systematically which techniques are best suited for determining various
quantities.

General analysis of surface brightness distributions is described in Section
2. The effects of stellar variability on observed IR emission are discussed in
Section 3. We analyse the observational implications of our results in Section
4 and summarize them in Section 5.

\section
                    {  SURFACE BRIGHTNESS DISTRIBUTION  }

The surface brightness of a spherically symmetric source depends only on the
impact parameter $b$, the displacement from the centre of symmetry.  In the
case of late-type stars, we have shown (IE95) that the solution of the
radiative transfer problem is a self-similar function of the scaled variable
$b/r_1$, where $r_1$ is the envelope's inner boundary, corresponding to the
dust condensation radius. For a given type of dust grains, this function is
virtually fully determined by the overall optical depth $\tau$ and does not
depend separately on luminosity, mass-loss rate, etc.

\begin{figure}

\caption{Normalized surface brightness distributions at 2.2, 5 and 10 \mic\
for models with dust condensation temperature $T_1$ = 800 K, stellar
temperature $T_\ast$ = 2500 K and optical depths as marked. Here $b$ and
$\theta$ are, respectively, the linear and angular displacements from the
centre, $r_1$ is the radius and $\theta_1$ is the angular diameter of the dust
condensation zone.}

\end{figure}

Fig. 1 presents the results of our model calculations at the three
representative wavelengths: 2.2, 5 and 10 \mic. The surface brightness profile
depends only on the dimensionless variable $b/r_1$, so that it scales
self-similarly with the dust condensation radius $r_1$. At any given
wavelength, the shape of this profile is determined by $\tau$; two different
systems with the same $\tau$ will have identical profiles in terms of $b/r_1$,
so that the actual value of $r_1$ is irrelevant.  For a system at distance $D$,
the impact parameter $b$ corresponds to an angular displacement $\theta = b/D$.
If we introduce the angular diameter of the dust condensation zone $\theta_1 =
2r_1/D$, then the observed surface brightness is a self-similar function of
$b/r_1 = 2\theta/\theta_1$.

In these calculations we have used a detailed model that takes into account
the wind acceleration by radiation pressure, determined self-consistently from
a solution of the coupled equations of hydrodynamics and radiative transfer
(for details see Netzer \& Elitzur 1993; IE95). Thanks to scaling, the
displayed profiles are applicable to any late-type star at the optical depths
indicated for each wavelength, with a given star corresponding to different
optical depths at different wavelengths. The models were calculated with the
opacities of amorphous carbon (Hanner 1988).  Other types of grains (silicate,
SiC, etc.) produce almost identical profiles for the same optical depths.  For
a given star, only the association of optical depths with wavelengths may
depend on grain chemistry when there are spectral features in the opacity. For
example, for the same optical depth at 2.2 \mic, a star with silicate grains
will have a larger 10-\mic\ optical depth than one with carbon grains. Thus
the two stars will have the same profile at 2.2 \mic\ but different ones at 10
\mic. The 10-\mic\ profile will be the same for the star with silicate grains
and another star with carbon grains whose $\tau$ is matched at that wavelength.

In addition to chemical composition, optical properties depend also on the
grain size $a$. Except for spectral features, the absorption and scattering
efficiencies are self-similar functions of the scaled variable $x = 2\upi
a/\lambda$ and their functional form does not change when $x < 1$.  Therefore,
so long as only $x < 1$ is covered by observations, the grain size is largely
irrelevant. For the wavelengths addressed here, $\lambda \ge 2.2$ \mic, all
grains smaller than \about\ 0.2 \mic\ are practically equivalent, and
following IE95 we adopt a single grain size of 0.05 \mic. The possible effect
of grains larger then 0.2 \mic\ is discussed below (Section 2.1). Other input
parameters used in these calculations are the stellar temperature \Tstar\ =
2500 K and the dust condensation temperature $T_1$ = 800 K.  The model results
depend primarily only on the ratio of these two temperatures. Varying $T_1$
and \Tstar\ within their relevant ranges has a negligible effect on the
results.

The narrow central component evident in the profiles is the stellar radiation
attenuated by the dust shell. Its width is $\rstar/r_1$, where \rstar\ is the
stellar radius.  This ratio is directly related to $\Tstar/T_1$.  The results
of our detailed model calculations, which account for the contributions of
both the central star and the dusty envelope to the radiation field at $r_1$,
can be adequately summarized with
 \eq{\label{r1/r*}
          { r_1 \over \rstar} = \alpha\left(\Tstar\over T_1 \right)^{\!2}.
 }
Here $\alpha$ is a dimensionless coefficient that depends primarily on the
grains' optical properties and only weakly on the overall optical depth. For
carbonaceous grains $\alpha$ is \about\ 1.2 for $\tau = 0$, increasing slowly
with $\tau$ to $\alpha \simeq 1.4$ at $\tau(2.2\ \mic) \sim 50$, the largest
plausible optical depth. For silicate grains, the corresponding range is
\about\ 0.5--0.8.  In addition, $\alpha$ has an even weaker dependence on
$T_1$ and \Tstar, which can be neglected in all practical applications.

The extended component visible in the figure is the contribution of the dust
shell.  At small $\tau$ this intensity component is proportional to the
optical depth along the line of sight, and the displayed profiles reflect the
behaviour of this function.  The features evident at $b/r_1 = \pm 1$ simply
reflect the envelope's inner edge.  High-resolution imaging will delineate
these features as a bright rim whose angular size ($\theta_1$) is wavelength
independent. The angular size of the dust formation zone is then directly
measurable.  In addition, the optical depth can be estimated by comparing the
measured ratio of peak intensities of the central (stellar) component and the
dust formation features with those of the displayed profiles.  As the optical
depth increases, these features gradually disappear and the shell's
contribution approaches that of the star.  The intensity profile becomes a
featureless bell-shaped curve, whose width is roughly proportional to
$\tau\theta_1$.  In this case it is impossible to determine $\tau$ and
$\theta_1$ independently.

\subsection
                           { Visibility curves }

Most spatially resolved IR observations of late-type stars are interferometric
and produce the two-dimensional Fourier transform of the surface brightness,
the visibility $V(q)$, where $q$ is the spatial frequency. For a circularly
symmetric intensity profile, $V(q) = F(q)/F(0)$, where
 \eq{\label{visibility}
                     F(q)= 2 \upi \int\limits_0^{\infty }
               I(\theta) J_0(2\upi q\theta)\, \theta \rm d\theta
 }
(cf.\ Rogers \& Martin 1984; $J_0$ is the Bessel function of zeroth order). In
the present case
 \eq{\label{vis-tot}
             V(q) = f_{\rm S} V_{\rm S} + f_{\rm D} V_{\rm D}\, ,
 }
where $V_{\rm S}$ and $V_{\rm D}$ are the individual visibility curves of the
stellar and dust emission, respectively, and $f_{\rm S}$ and $f_{\rm D}$ are
the fractional contributions of these two components to the total flux.  The
stellar component can be closely approximated by a uniform-brightness disc
 \eq{\label{vis-star}
      V_{\rm S} = {{2 J_1(\upi q\theta_{\ast})} \over {\upi q\theta_{\ast}}},
 }
where $J_1$ is the Bessel function of first order and $\theta_{\ast} =
2\rstar/D$ is the angular diameter of the stellar disc.  The stellar
visibility is a function of $q\theta_{\ast}$.  Because of scaling, the shell's
contribution to the intensity is a self-similar function of $\theta/\theta_1$,
thus the corresponding visibility $V_{\rm D}$ is a self-similar function of
$q\theta_1$.

\begin{figure}

\caption{Visibility curves ($q$ is the spatial frequency) at 2.2, 5 and 10
\mic\ for models with the same temperatures as in Fig.~1 and optical depths
as marked.  The dashed line is the visibility function of the naked star
($\tau = 0$).}

\end{figure}

Fig.~2 presents the visibility curves of the surface brightness
distributions displayed in Fig.~1, including more values of $\tau$ for
optically thin envelopes. Just as the surface brightness profiles scale
self-similarly with $r_1$, the visibility profiles scale self-similarly with
$\theta_1$. The dashed lines are the visibility curves of a naked star,
$V_{\rm S}$, included for comparison. They are quite flat, reaching zero
visibility only at $q\theta_1 = 1.2\, \theta_1/\theta_{\ast} = 1.2\,
r_1/\rstar$. Note that the ratio $r_1/\rstar$, typically \about\ 5--10, is
directly related to the temperature ratio $T_1/\Tstar$ (equation~\ref{r1/r*}).

The stellar and envelope contributions are easily discernible in the figure.
All visibility curves display an initial drop with $q\theta_1$, reflecting the
extended nature of the dust component, followed by a levelling off when the
envelope is fully resolved and the stellar component takes over.  The value
$V_{\rm c}$ of the visibility at that levelling-off is roughly the fraction of
the stellar contribution to the total flux.  The significance of this
contribution depends on both optical depth and wavelength. When it is
appreciable, $V_{\rm c}$ is finite and the levelling-off occurs at $q\theta_1
\sim 1$. The visibility curve can then be used to determine both $\theta_1$
and $\tau$ -- the magnitude of $V_{\rm c}$ directly gives $\tau$, and the
location of the levelling-off determines the angular scale $\theta_1 \simeq
1/q$. In sources where $V_{\rm c} \simeq 0$, the levelling-off occurs at
progressively smaller $q\theta_1$ as $\tau$ increases, a direct consequence of
the broadening of the intensity profile (cf.\ Fig.~1). The visibility curves
become self-similar with angular size-scale roughly proportional to $\tau
\theta_1$, thus $\tau$ and $\theta_1$ cannot be determined independently.
These points have been recognized previously in the discussion of surface
brightness.

It is important to note that the dust component of the visibility profiles
always varies with wavelength, even in optically thin envelopes whose images
are dominated by wavelength-independent bright rims (corresponding to the dust
formation zone). The reason is that, at any given wavelength, the visibility
reflects the angular size of the region from which most of the flux is
emitted. This size corresponds to the radius where the dust temperature is
comparable to the Wien temperature of that wavelength. Since the dust
temperature decreases with distance, the size of the flux emitting region
increases with wavelength. Such variation has been detected in many
observations. Most visibility observations are interpreted in terms of
Gaussian fits and are reported in terms of the resulting Gaussian FWHM
$\theta_{\rm G}$. A quantitative comparison of theory with observations must
incorporate this widespread method of reporting observational results. The
observed Gaussian widths are found to increase with wavelength according to
$\theta_{\rm G} \propto \lambda ^k$, and the index $k$ is found to vary
considerably among individual stars. In the wavelength range 2--10 \mic, Dyck
et al.\ (1984) obtained $k$ \about\ 0.2 for C stars (carbon dust) and $k$
\about\ 0.7--1.3 for O stars (silicate dust).  Cobb \& Fix (1987) found $k
\sim 1$ in the wavelength range 5--10 \mic\ for a sample of OH/IR stars
(silicate dust).

Our models readily explain these detailed observations. Fitting a Gaussian to
the model visibilities, we find that in the wavelength range 2.2--10 \mic\
these fits can be summarized with
\eq{\label{thetaG}
   \theta_{\rm G} \simeq \theta_1\left({\lambda \over 3.2\,\mic}\right)^{\!0.6}
     \times\cases{1             & $\tau \le 1$   \cr
                                                 \cr
                  \tau^{0.4}    & $\tau >   1$,  \cr }
}
an approximation accurate to within 10--20 per cent. Therefore, in optically
thin envelopes $k$ = 0.6 independent of grain chemical composition, in
agreement with the lower bound of the range found in O stars by Dyck et al. In
optically thick envelopes, on the other hand, $k$ does vary with the grain
chemistry because of its dependence on $\tau$.  We find that $\theta_{\rm G}
\propto \lambda^{0.2}$ in amorphous carbon optically thick envelopes, in
agreement with the observations of C stars.  The results for silicate grain
optically thick envelopes are more involved because of the peak at 9.7 \mic.
A parametrization in the form $\theta_{\rm G} \propto \lambda^k$ results in
$k$ = 1.3 when the fit applies to the wavelength range 2--10 \mic\ and $k$ = 4
for the range 7--10 \mic.  These results are in agreement with the various O
star observations quoted above.

In addition, observations by Fix \& Cobb (1988) at 8 and 10 \mic\ show that
$\theta_{\rm G}$ is roughly the same at both wavelengths in IRC+10420 but
increases by about a factor of 2.5 in OH 26.5+0.6, even though both stars
display the silicate feature.  Our results explain both observations if the
dust shell is optically thin in IRC+10420 (expected $\theta_{\rm G}$ variation
of only 10 per cent) and thick in OH 26.5+0.6 (expected variation by a factor
of 2.4).  Indeed, the former displays the silicate feature in emission, the
latter in absorption, indicating optically thin and thick envelopes,
respectively.

In spite of these successes, our model results indicate that single-Gaussian
fits to visibility curves must be used with great care.  Such fits are
adequate within observational uncertainties (\about\ 0.05 in visibility) only
at short wavelengths ($\la$ 2.2 \mic).  The quality of Gaussian fitting
deteriorates as the wavelength increases, and we find such fitting meaningless
for $\lambda \ga$ 10 \mic. The apparent success of Gaussian fits at such
wavelengths in a number of actual observations usually reflects partial
coverage of the visibility curve (e.g.\ Sutton, Betz \& Storey 1979,
$V \le 0.5$; Cobb \& Fix 1987, $V \ge 0.8$). Indeed, when the spatial frequency
coverage provides the entire visibility curve, single-Gaussian fits become
inadequate (e.g.\ Mariotti et al.\ 1983; Dyck et al.\ 1987). While such a
fitting procedure fails, our models provide satisfactory fits to all observations
and provide a more adequate method for data analysis.

Finally, for given a grain size $a$, scattering is important only for
wavelengths shorter than $\lsca = 2\upi a$. For the size used here of 0.05
\mic, \lsca\ is 0.3 \mic\ and scattering is negligible at all the displayed
wavelengths.  For any plausible grain size ($\la$ 1 \mic), scattering can
become important only at the shortest displayed wavelength of 2.2 \mic.  When
that happens, the shape of the visibility curves remains essentially the same
but the corresponding optical depth is significantly smaller. For example, if
the grain size is 0.2 \mic, the five curves in the topmost panel of Fig.~2 are
produced with optical depths of 0.1, 0.8, 1, 3 and 10, respectively, instead
of the indicated values of 0.1, 1, 3, 10 and 30.

\section                    { TIME VARIABILITY }

Late-type stars display variability with periods of order years and bolometric
amplitudes of 1--2 mag (e.g., Blommaert 1992). The bolometric luminosity
variability affects the infrared properties through changes of the dust
condensation radius and overall optical depth.  Combining equation
(\ref{r1/r*}) with the luminosity relation $L_\ast = 4\upi r_\ast^2\sigma
T_\ast^4$ produces
 \eq{\label{r1}
   r_1 =  2.3\times10^{14} \alpha\left({L_\ast \over 10^4\ \Lo }\right)^{\!1/2}
               \left({ 10^3 {\rm \ K}} \over T_1 \right)^{\!2}\ {\rm cm}.
 }
A similar relation for galactic nuclei was noted in the recent numerical work
of Laor \& Draine (1993). Luminosity variability induces movement of the dust
condensation point: the shell's inner boundary is furthest from the star at
maximum light.  This variation in linear scale-size of the envelope translates
directly to its angular scale size, $\theta_1$. {\em Envelopes are larger
during maximum luminosity}. It is noteworthy that variations of stellar radius
and temperature during the stellar cycle do not enter independently, only
through the overall luminosity.

Such dependence of envelope size on phase of the light-curve has been
detected. McCarthy, Howell \& Low (1978) found that Mira ($o$ Ceti) was
partially resolved at 10.2 \mic\ near maximum visual phase but unresolved at
minimum phase. McCarthy, Howell \& Low (1980) measured the size of IRC+10216
at 5 \mic\ and obtained larger value at maximum than at minimum of the
light-curve. Danchi et al.\ (1990, 1994) reobserved this source at 11 \mic\
and confirmed these results. In a separate study (Ivezi\'{c} \& Elitzur 1996;
hereafter IE96) we analyse in detail all observations of IRC+10216 and show
that the angular diameter of its dust condensation zone varies between \about\
0.3 and \about\ 0.5 arcsec as the luminosity changes from minimum to maximum.

Variations of $r_1$ during the stellar cycle affect the dust optical depth so
that $\tau$ should be smallest at maximum light, largest at minimum. If the
grain optical properties do not change during the stellar cycle, the overall
optical depth is proportional to the dust column density. For constant
mass-loss rate \Mdot\ and velocity $v_1$ at the base of the envelope, the dust
column density scales with $r_1^{-1}$.  Together with equation (\ref{r1}),
this implies $\tau \propto L_\ast^{-1/2}$.  Since both \Mdot\ and $v_1$ may
vary during the stellar cycle, the dependence of $\tau$ on $L_\ast$ can be
expected to be somewhat different but still show the same trend.  From
detailed analysis we find that, in IRC+10216, $\tau \propto L_\ast^{-0.2}$
(IE96). Because of the variation of $\tau$, IR spectral properties should vary
with the light curve, corresponding to redder radiation at minimum light
(largest optical depth), bluer at maximum. Indeed, ever since the work of
Pettit \& Nicholson (1933), such behaviour has manifested itself in many
observational correlations. In particular, Le Bertre (1988a,b) finds that in
several dusty envelopes the wavelength of peak emission is shorter at maximum
than at minimum light.  In general, {\em envelopes are bluer around maximum
light}.

The phase dependence of optical depth can be analysed quantitatively with the
aid of photometric amplitudes $A = 2.5 \log (F^{\rm max}/F^{\rm min})$, where
$F^{\rm max}$ and $F^{\rm min}$ are fluxes at maximum and minimum light,
respectively.  At each wavelength, the photometric amplitude $A_\lambda$ is
related to the amplitude $A_{\rm bol}$ of the bolometric flux $F$ via
 \eq{\label{Alambda}
                   A_\lambda = A_{\rm bol} + a_\lambda\,,
 }
where $a_\lambda$ is the amplitude of normalized flux $f_\lambda =
F_\lambda/F$.  Because of scaling, $f_\lambda$ does not depend separately on
luminosity, mass-loss rate, etc., only on overall optical depth (IE95).
Therefore, $a_\lambda$ is a unique function of $\tau$, specifically
\eq{\label{Af}
    a_\lambda(\tau) = 2.5\log{f_\lambda(\tau^{\rm max}) \over
                              f_\lambda(\tau^{\rm min})},
}
where $\tau^{\rm max}$ and $\tau^{\rm min}$ are the optical depths at maximum
and minimum light, respectively.  That is, for a given dust chemical
composition, $a_\lambda$ too does not depend on individual properties other
than optical depth. Our detailed model calculations show that \alam\ has a
simple behaviour in two distinct limits.  At wavelengths for which $\tau < 1$,
typically $\lambda \ga$ 20 \mic, $\alam \propto -\Abol$. At wavelengths for
which emission can be neglected, typically $\lambda \la$ 3 \mic\ (wavelengths
shorter than the peak of the Planck function of the dust formation temperature
$T_1$), $\alam \propto \tau_\lambda$. From these results we find that
 \eq{
  A_\lambda \simeq \cases{\Abol + 0.7\tau_\lambda & $\lambda \la 3\ \mic$  \cr
                                                                           \cr
                         (0.5-0.7)\times\Abol     & $\lambda \ga 20\ \mic$.\cr}
}
The photometric amplitudes vary linearly with $\tau$ in the near-IR, becoming
wavelength independent in the mid- and far-IR.  Since \Abol\ varies little
among stars while $\tau$ varies a lot, the range of photometric amplitudes can
be expected to be large at short wavelengths, small at long ones. In addition,
$\tau$ decreases with wavelength in the near-IR, thus the amplitudes should
decrease too.

These simple results readily explain the available observations.
Short-wavelength amplitudes indeed decrease with wavelength and visual
amplitudes are usually much larger than bolometric ones (e.g.\ Hetzler 1936;
Lockwood \& Wing 1971). Furthermore, analysis of the Harvey et al.\ (1974)
observations shows that near-IR amplitudes of different stars increase with
overall optical depth, and for a given star decrease with wavelength. This
behaviour is confirmed by subsequent observations (e.g. DiGiacomo et al.\
1991). Finally, the spread among stars is much larger for visual than for
bolometric amplitudes (e.g.\ Forrest, Gillet \& Stein 1975). For long
wavelengths, Herman et al.\ (1984) find that the variation of OH maser
luminosity, and presumably far-IR flux at 35 \mic, is about half that of the
bolometric luminosity.

Dust spectral features can display a more complex variability with the stellar
phase. Depending on initial optical depth, the strength of the 9.7-\mic\
silicate emission feature can either increase or decrease for the same change
of $\tau$: in optically thin envelopes the strength increases with $\tau$
until it reaches maximum at some optical depth $\tau_{\rm m}$ (\about\ 1),
after which it decreases because of self-absorption (IE95).  As a result, in
envelopes with $\tau < \tau_{\rm m}$ the 9.7-\mic\ emission feature is
strongest during minimum light (the optical depth is largest), with the
reverse behaviour in envelopes with $\tau > \tau_{\rm m}$. In envelopes with
$\tau \sim \tau_{\rm m}$, the emission feature can remain almost unchanged
during the stellar cycle. Little-Marenin, Staley \& Stencel (1993) find
examples for each type of behaviour in a number of sources that display the
9.7-\mic\ emission feature. On the other hand, when this feature is in
absorption, its depth always increases with $\tau$ (Scoville \& Kwan 1976;
IE95).  Thus it is always deepest at minimum light, as observed (e.g. Engels et
al.\ 1983).

\section {            OBSERVATIONAL IMPLICATIONS              }

The source intensity is a unique function of various intrinsic properties such
as luminosity, mass-loss rate, dust density, dust-to-gas ratio, etc.  However,
because of scaling, intrinsic properties do not enter separately, only
indirectly through their effect in determining $\tau$ and $r_1$.  For given
grain optical properties, the surface brightness is a unique function of the
scaled variable $b/r_1$, essentially determined by overall optical depth.
Since optical depths involve only intrinsic properties, the same applies to
the self-similar profiles displayed in Figs 1 and 2. Contact with observations
brings in the distance to the source, which fixes the scales of angular size
$\theta_1$ and bolometric flux $F$; that is, the distance sets the scale of
conversion from linear to angular sizes and from intrinsic to observed
fluxes.  An additional dependence on the intrinsic parameters $T_1$ and
\Tstar\ does exist, but it is rather weak and affects only optically thin
envelopes at short ($\la$ 7 \mic) wavelengths.  Furthermore, compared with the
variation range of $\tau$ and $\theta_1$, $T_1$ and \Tstar\ are practically
constant among late-type stars. As a result, the only non-radiative system
properties that generally can be determined by the most detailed IR
observations are $\tau$ (including its wavelength dependence), a purely
intrinsic property, and $\theta_1$, which sets the scales for observations. We
discuss now which observational techniques are most suitable for determining
these two quantities.

The least detailed observations are unresolved flux measurements, determining
the spectral energy distribution $F_\lambda$.  This function is the product of
the spectral shape $f_\lambda$ and the bolometric flux $F$. For given grains,
the intrinsic property $f_\lambda$ is controlled by $\tau$ and is independent
of $\theta_1$, thus it cannot be used to determine $\theta_1$.  However, for
this very reason -- its dependence on a single quantity -- spectral shape
analysis is the most reliable method for determining $\tau$ and its wavelength
dependence, resulting in the best handle on the dust chemical composition. The
additional information contained in the bolometric flux allows an estimate of
$\theta_1$, described below, albeit an indirect one.

Direct determination of $\theta_1$ requires spatially resolved observations.
However, as shown in Section 2, $\theta_1$ and $\tau$ enter independently only
in optically thin envelopes.  In such sources, $\theta_1$ can be directly read
off the radial profiles from the position of the dust features, and $\tau$
determined from the intensity ratio of stellar and dust components (cf.\ Fig.\
1).  This determination of $\tau$ is considerably less certain than that of
$\theta_1$. As a result, the surface brightness distribution is always a poor
indicator of $\tau$, suggesting that the most practical approach involves a
two-step data analysis procedure. In the first step $\tau$ is determined from
a fit to the spectral shape, and in the second the resulting model prediction
for the surface brightness is used to determine $\theta_1$.  We have used such
a two-step approach to analyse the IR observations of IRC+10216, and with a
single model obtained successful fits to all spectral and spatially resolved
observations (IE96).

We proceed now to a more detailed discussion of the capabilities of different
high-resolution techniques.

\subsection                { Direct imaging }

Direct imaging is the best method for determining $\theta_1$, from the
location of the features corresponding to dust condensation. Unfortunately,
these features are visible only for a limited range of optical depths.  At
large $\tau$ the features are smeared out as the profile evolves into a bell
shape. At small values of $\tau$, the features are so weak compared with the
stellar component that they become undetectable at the dynamic range of most
current telescopes; note, however, that on telescopes such as {\it HST}, Keck
or Gemini, a dynamic range of 1000 is possible in images at spatial
resolutions better than 0.1 arcsec. As a result, at every wavelength the
features are a useful indicator only when the optical depth is of order unity.
Thus the features generally cannot be used at either end of the spectrum: at
short wavelengths ($\la$ 2--3 \mic), the dust contribution is minute in
comparison with the stellar component when $\tau \sim 1$ so the features are
hard to resolve; and, at long wavelengths ($\ga$ 20--30 \mic), $\tau$ never
reaches unity for the majority of late-type stars (IE95). Therefore, {\em the
mid-infrared is the best spectral range for resolving the dust condensation
features in surface brightness profiles}.

\begin{table*}
 \begin{minipage}{6in}
 \begin{center}

\caption{Best candidates for imaging: Late-type stars identified from their
{\it IRAS} fluxes as the most promising candidates for potential resolution of
the dust condensation zone in future observations.  Quantities tabulated in
columns (d)--(f) are model-based estimates.}

\smallskip
\begin{tabular}{@{}rlcrccc@{}}
\hline
   {\it IRAS} Name & Other Name\n(a) & LRS\n(b) & $F_{12}$\n(c)
   & $\tau_{\rm F}$\n(d) & $\tau_{10}$\n(e) & $\theta_1$\n(f)   \\
\noalign{\smallskip}
 05027$-$2158  & T Lep    & 15 &   157 &  1.36 & 0.23  &  0.03 \\
 11294$-$6257  & na       & 15 &     5 &  3.95 & 1.76  &  0.05 \\
 12380+5607    & Y UMa    & 15 &   193 &  0.89 & 0.15  &  0.04 \\
 17297+1747    & V833 Her & 14 &   559 &  2.24 & 0.38  &  0.05 \\
 20120$-$4433  & RZ Sgr   & 16 &    38 &  1.91 & 0.32  &  0.02 \\
 03507+1115    & IK Tau   & 26 &  4630 &  0.12 & 0.39  &  0.22 \\
 04566+5606    & TX Cam   & 27 &  1640 &  0.05 & 0.17  &  0.22 \\
 07209$-$2540  & VY CMa   & 24 &  9920 &  0.27 & 0.87  &  0.31 \\
 18050$-$2213  & VX Sgr   & 26 &  2740 &  0.11 & 0.37  &  0.20 \\
 23558+5106    & R Cas    & 24 &  1340 &  0.06 & 0.20  &  0.17 \\
 09116$-$2439  & na       & 42 &   737 &  1.43 & 0.22  &  0.07 \\
 15082$-$4808  & na       & 42 &   793 &  1.41 & 0.21  &  0.08 \\
 17049$-$2440  & na       & 42 &   793 &  1.79 & 0.27  &  0.08 \\
 18240+2326    & na       & 42 &   731 &  1.73 & 0.26  &  0.07 \\
 23320+4316    & LP And   & 42 &   959 &  1.09 & 0.16  &  0.09 \\ \hline

\end{tabular}
\end{center}
\smallskip

(a) na: {\it IRAS} source without known association \\
(b) {\it IRAS} LRS class\\
(c) {\it IRAS} flux at 12 \mic\ (Jy) \\
(d) Flux-averaged optical depth \\
(e) Optical depth at 10 \mic \\
(f) Angular size of the dust condensation zone (arcsec) \\

\end{minipage}
\end{table*}

Table 1 lists the late-type stars we have identified as the best candidates for
observations that can potentially resolve the dust formation features. They
were selected as the stars with the largest expected values of $\theta_1$ among
{\it IRAS} sources whose 10 \mic\ optical depth is in the range 0.1--2.  In
searching the {\it IRAS} catalogue we were careful to include only sources with
the highest flux qualities in all four wavebands and uncontaminated by cirrus
emission (see IE95 for details of the selection process and removal of
contaminated sources).  The optical depth was determined from the {\it IRAS}
fluxes through extensions of the techniques of IE95, described in the Appendix.
Estimating the angular size $\theta_1$ is more involved.  Our starting point is
the relation
\eq{\label{theta1}
     \theta_1 = 0.17 \, \alpha
      \left(F \over 10^{-8} {\rm\,W\,m^{-2}}\right)^{\!0.5}
      \left({ 10^3 {\rm \,K}}\over T_1 \right)^{\!2} \hbox{ arcsec},
}
obtained by combining equation (\ref{r1}) with $L_\ast=4\upi D^2F$; note that
when $\theta_1$ and $F$ are directly measured in observations, this equation
can be inverted to determine the dust condensation temperature $T_1$. The
Appendix describes a method for estimating the bolometric flux $F$ from the
{\it IRAS} fluxes, and for simplicity we assume $T_1$ = 800 K for all stars.
The resulting estimates for $\theta_1$ are expected to be accurate to within a
factor of 2 or so. Table 1 presents the five sources with the largest
$\theta_1$ that we identified in each of the following LRS classes: 10
(featureless spectra), 20 (silicate 9.7-\mic\ feature in emission) and 40
(11.3-\mic\ in emission, mixture of amorphous carbon and SiC grains).

\subsection      {     Visibility observations    }

All the radiative information available is contained in the surface brightness
distribution. Therefore, direct imaging would be in principle the ultimate
method to determine the system properties.  However, finite dynamical range
limitations preclude direct imaging of systems with very thin envelopes, where
the intensity of dust emission is negligible in comparison with the stellar
component.  The structure of such systems can still be probed by visibility
observations for the following reason. As is evident from equation
(\ref{visibility}), visibility is a measure of flux rather than intensity. And
because of the much larger area of the dust emission, its flux can exceed that
of the star even when the relation between intensities is the
reverse.\footnote{This also implies that the stellar component can dominate
the intensity profile even when it is negligible in the visibility
observations. This is contrary to a conclusion by Griffin (1990).}

At wavelengths shorter than \about\ 2 \mic\ and albedo $\ga 0.1$, the emission
component of the dust radiation field is much weaker than either the scattered
component or direct stellar contribution because it requires dust warmer than
plausible condensation temperatures.  The radiation field is mostly composed
of the attenuated stellar component and scattered light, and we find from our
detailed solutions that under these circumstances
\eq{
                  V_{\rm c} \simeq \exp(-{\tau_{\rm sca}}),
}
where $\tau_{\rm sca}$ is the scattering optical depth. This relation has a
simple physical explanation and it is valid for $\tau_{\rm sca} \la 2$.  With
this relation, visibility observations can determine the wavelength dependence
of $\tau_{\rm sca}$, a variation that can then be used to estimate the grain
size $a$.  The scattering optical depth is constant at short wavelengths,
changing its behaviour to $\tau_{\rm sca} \propto \lambda^{-4}$ for $\lambda
\ga \lambda_{\rm sca} = 2\upi a$. Determining $\lambda_{\rm sca}$ from the
change in behaviour of $\tau_{\rm sca}$ yields an estimate for $a$. We have
applied this method to the visibility observations of IRC+10216 and obtained
$a \sim$ 0.2 \mic, in agreement with an independent estimate based on spectral
shape analysis (IE96).

\section
                               {DISCUSSION}

As shown in our previous study (IE95), the dynamics and spectral energy
distributions of dusty winds around late-type stars are controlled by the dust
chemical composition and overall optical depth. Here we have shown that a
complete description of observations requires one additional parameter -- the
angular size of the dust condensation zone. Since the spectral shape is
entirely independent of $\theta_1$ it is the most reliable indicator of
$\tau$, whereas spatially resolved observations provide the most useful
measurement of the dust condensation zone. Thus the most practical method of
analysis is to determine the optical depth from model fits to the spectral
shape, and $\theta_1$ from high-resolution observations (either direct imaging
or visibility). Because of scaling, the model results presented here and in
IE95 provide sufficient coverage of parameter space for the analysis of
virtually any late-type star.  An example of such complete analysis is
presented in IE96 for IRC+10216.

Our models do not include the effects of light travel time between the central
star and the heated dust.  The phase lags that such travel introduces between
variations of the dust and stellar components are insignificant at short- and
mid-infrared wavelengths, whose emission originates close to the dust
condensation zone.  At long wavelengths, however, the dust emission originates
sufficiently far from the star that the phase lags can become appreciable.
Analysis of this effect has been recently reported by Wright \& Baganoff
(1995), who show that it can be used to determine the distance to the star. It
is important to note that these effects do not invalidate our steady-state
modelling, as discussed in IE95.

In optically thin envelopes, our model images have a bright rim at the dust
condensation radius, leading to the only features, except for the stellar
spike, in the surface brightness distribution. Images of optically thick
envelopes do not display any details. We did not consider here the interaction
of the wind with the surrounding interstellar medium, which may produce a
dense shell far from the star (IE95). Such a shell would produce additional
features in the surface brightness profile at long wavelengths ($\ga$ 50
\mic), which may have been observed for some {\it IRAS} stars at 60 and 100
\mic\ (Young, Phillips \& Knapp 1993). Indeed, from the observed angular sizes
of these features Young et al.\ conclude that they probably correspond to the
interface of the wind and the interstellar medium.

\section*{Acknowledgments}

This research has made use of the {\sc SIMBAD} database, operated at CDS,
Strasbourg, France, and the ADS database. Support by NSF grant AST-9321847,
NASA grant NAG 5-3010 and the centre for Computational Sciences of the
University of Kentucky is gratefully acknowledged.

\appendix

\section   { OPTICAL DEPTHS AND BOLOMETRIC FLUX FROM {\it IRAS} DATA }

Dust emission from late-type stars with envelopes is controlled by a single
parameter, the flux averaged optical depth $\tau_{\rm F}$.  The dependence of
various spectral quantities on $\tau_{\rm F}$ is listed by IE95, who also
describe a method for determining $\tau_{\rm F}$ from {\it IRAS} fluxes.  Here
we provide some additional useful relations, derived from our full model
calculations.

Once $\tau_{\rm F}$ is known, the optical depth at any wavelength can be
obtained from
\eq{
                     \tau_\lambda = t_\lambda \tau_{\rm F}^n.
}
The power $n$ is 1 for all dust grains when $\tau_{\rm F} \le 1$.  When
$\tau_{\rm F} > 1$, $n$ is 0.5 for silicate-based grains and 1.7 for
carbonaceous grains. The constants $t_\lambda$ at selected wavelengths are
listed in Table A1 for various grain compositions.  For other wavelengths,
$t_\lambda$ can be interpolated using fig.\ 1 of IE95.  Extrapolation toward
shorter wavelengths is risky because of the increased contribution of
scattering.

An additional useful correlation involves the bolometric flux and {\it IRAS}
fluxes:
\eq{\label{Fbol}
      F = C\times10^{-12} F_{60}
 \left( {F_{12} \over F_{25}} \right) ^{\gamma} \quad {\rm W\,m^{-2}}.
}
Here $F_{12}$, $F_{25}$ and $F_{60}$ are {\it IRAS} fluxes in Jy, and $C$ and
$\gamma$ are constants that depend on the dust chemical composition. For
silicate-based grains $C=7$ and $\gamma = 3$.  For carbonaceous grains
$C=0.08$ and $\gamma = 6$ when $F_{12} > 2 F_{25}$ and $C=1.8$, $\gamma = 1.5$
otherwise.

\begin{table}
  \begin{center}
  \caption{Constants for equation (A1).}
  \begin{tabular}{@{}ccccc@{}}
   \hline
   $\lambda$ (\mic) & Sil.\n(a) & Oli.\n(b) & am.C\n(c) & SiC\n(d) \\
   \noalign{\smallskip}
            \02.2 & 0.4 & 0.4 & 1.20 & 1.10  \\
             10.0 & 3.2 & 3.6 & 0.17 & 0.15  \\
             12.0 & 1.5 & 1.9 & 0.12 & 0.18  \\
             25.0 & 1.2 & 1.5 & 0.05 & 0.04  \\
             60.0 & 0.2 & 0.2 & 0.01 & 0.01  \\ \hline

\end{tabular}
\end{center}

\smallskip

(a) Astronomical silicate \\
(b) Mixture of 80 per cent (by mass) astronomical silicate and 20 per cent
    crystalline olivine \\
(c) Amorphous carbon \\
(d) Mixture of 80 per cent (by mass) amorphous carbon and 20 per cent SiC \\

\end{table}

\label{lastpage}

\end{document}